\documentclass[aps,prapplied,reprint,author-year,showpacs,longbibliography]{revtex4-2}

\usepackage{amssymb}
\usepackage{lipsum}
\usepackage{color}
\usepackage{graphicx}
\usepackage{amsmath}

\newcommand{\JUAddress}{Institute of Physics, Jagiellonian University in Krak\'ow, \L{}ojasiewicza 11, 30-348 Krak\'ow, Poland} 

\begin{document}

\title{Rotating polarization magnetometry}

\author{Szymon Pustelny}
\affiliation{\JUAddress}
\affiliation{Department of Physics, Harvard University, Cambridge, MA 02138, USA}
\email{pustelny@uj.edu.pl}

\author{Przemys\l{}aw W\l{}odarczyk}
\affiliation{\JUAddress}

\begin{abstract}
Precise magnetometry is vital in numerous scientific and technological applications. At the forefront of sensitivity, optical atomic magnetometry, particularly techniques utilizing nonlinear magneto-optical rotation (NMOR), enables ultraprecise measurements across a broad field range. Despite their potential, these techniques reportedly lose sensitivity at higher magnetic fields, which is attributed to the alignment-to-orientation conversion (AOC) process. In our study, we utilize light with continuously rotating linear polarization to avoid AOC, producing robust optical signals and achieving high magnetometric sensitivity over a dynamic range nearly three times greater than Earth's magnetic field. We demonstrate that employing rotating polarization surpasses other NMOR techniques that use modulated light. Our findings also indicate that the previously observed signal deterioration is not due to AOC, suggesting an alternative cause for this decline.
\end{abstract}

\maketitle

\section{Introduction}
\label{introduction}

Optical magnetometry belongs to one of the most mature quantum technologies to date  \cite{Budker2007Optical,Budker2013Optical,Budker2000NMOR,Allred2002High,Alexandrov2003Recent,Seltzer2007Synchronous,Groeger2006High,Pustelny2008Magnetometry,Kim2016Ultra,Colombo2016Four,Oelsner2019Performance,Bevington2019Imaging,Wang2020Combined,Fang2020High}. Modern optical atomic magnetometers (OAMs) offer a magnetic-field sensitivity often exceeding 10~fT/Hz$^{1/2}$, with the only sensitivity-wise competitors being bulky and high-maintenance superconducting quantum interference devices (SQUIDs). This remarkable sensitivity, combined with technical simplicity and low operating costs, has led to a wide range of applications, from biomagnetic field measurements \cite{Bison2003Magnetocardiography,Murzin2020Ultrasensitive,Brooks2022Magnetoencephalography} and surveys natural resources \cite{Lu2023Geophysics,Akulshin2025Remote} to zero-field nuclear magnetic resonance detection \cite{Bevilacqua2016Microtesla,Tayler2017ZULFReview,Barskiy2025ZULF} and searches for ultra-light dark matter \cite{Pustelny2013GNOME,Jiang2021Search,Terrano2022Comagnetometry,Afach2024What,Bloch2024Curl} (for more information, see, for example, Ref.~\cite{Budker2013Optical} and references therein).

The highest sensitivity of OAMs is achievable at very weak magnetic fields, necessitating effective shielding against external and uncontrolled magnetic fields. Operation at stronger fields, such as Earth's magnetic field, requires the implementation of solutions that can, at least partially, capitalize on the low-field sensitivity. For example, the stronger field could be compensated using a set of calibrated magnetic-field coils to bring the operating point of the magnetometer back to zero \cite{Seltzer2004Unshielded,Belfi2007Cesium,Chu2014Active,Papoyan2016Magnetic}. By actively bringing the field to zero, one can (nominally) maintain the zero-field sensitivity, additionally obtaining directional sensitivity through magnetic-field modulation \cite{Seltzer2004Unshielded,Papoyan2016Magnetic}. Unfortunately, due to the limited stability of current sources and inhomogeneities induced by compensation of the magnetic fields, these methods still suffers from a sensitivity loss. 

An alternative technique that has recently regained interest is based on the observation of magnetization decay of optically polarized atomic vapors \cite{Limes2020Portable,Wilson2020Wide,Lucivero2022Femtotesla,Zhang2022Multipass,Hunter2023Field}. By analyzing a time-dependent signal arising after the pulsed polarization of the atoms, one determines the spin precession frequency and hence the magnetic field to which the atoms are exposed to. Despite its utility, this approach does have some drawbacks, including (in some cases) the relatively involved signal analysis and inability to perform continuous measurements.


The third group of methods is based on modulating certain physical parameters of the system to enhance sensitivity in high fields. A notable example of this approach is the so-called M$_x$ magnetometry \cite{Alexandrov1996Double,Groeger2006High,Scholtes2016Suppression}, which is based on the application of an external, oscillating magnetic field and observation of light transmission through the magneto-optically active medium. By an appropriate choice of the oscillating-field frequency, one can resonantly excite spin precession, generating the strongest dynamic response of the medium. In turn, by tracking the resonance position, one can monitor the magnetic-field strength. While in the M$_x$ magnetometers the circularly polarized light is typically used, similar measurements can be achieved with linearly polarized light \cite{Ledbetter2007Detection,Put2019Nonlinear}. In this case, one can either detect static or oscillating magnetic fields by controlling the other field frequency or magnitude \cite{Put2019Different}. The drawback of these techniques, however, is the necessity of applying an external field, which, in the case of multiple sensors, for example, for biomagnetic field imaging \cite{Bison2003Magnetocardiography}, leads to cross-talk between the devices. Alternatively, the additional field can be replaced by modulation of interacting-light parameters \cite{Budker2013Optical}. This was already demonstrated in the first Bell-Bloom magnetometer \cite{Bell1961Optically}, where the pumping light intensity was modulated, giving rise to the so-called high-frequency resonance arising when the modulation frequency coincides with the magnetization-precession (Larmor) frequency of atoms. This technique was further developed in the scope of magnetometric techniques exploring nonlinear magneto-optical rotation (NMOR) \cite{Budker2000NMOR}, where both frequency \cite{Budker2002FMNMOR,Acosta2006Geomagnetic} and amplitude \cite{Gawlik2006Nonlinear,Pustelny2008Magnetometry} modulation were used to measure stronger magnetic fields. 

A problematic feature of NMOR magnetometers is the deterioration of the amplitude of high-frequency resonances \cite{Acosta2008Production}, and hence the magnetic-field sensitivity, with increasing magnetic field. It is known that this deterioration and broadening of the resonance can be attributed to the splitting observed in the NMOR resonance due to the nonlinear Zeeman effect, where resonances associated with different magnetic sublevels are unequally shifted \cite{Acosta2006Geomagnetic,Chalupczak2010Competition}. Additionally, it has been postulated that this effect may also originate from the so-called alignment-to-orientation conversion (AOC) \cite{Budker2000AOC}. This effect arises when the electric field of light is not parallel to the atomic polarization and it consists of the transfer of the atomic polarization (alignment) that is detectable in NMOR into atomic polarization (orientation) that is undetectable with the method. It should be noted, however, that the process of AOC conversion arises when there is a non-zero angle between light and atomic polarizations, i.e., it disappears when they are parallel.

In this work, we present the application of light with rotating linear polarization for optical magnetometry. By precisely matching the atomic Larmor frequency with the frequency of polarization rotation, we generate dynamic polarization of the medium. This dynamic atomic polarization leads to the polarization rotation of the second initially unmodulated light beam. The parameters of the induced NMOR resonances are measured against factors such as light intensity and tuning. Since, the orientation of atoms and light polarization remains parallel at all times, i.e., AOC does not arise, this arrangement allows us to test the role of the effect in the deterioration of magnetometric sensitivity at higher fields. Specifically, we conduct a comparative analysis between the rotating polarization (RotPol) and amplitude-modulated NMOR (AMOR) methods, as AOC should play a detrimental role in the latter case. Through these investigations, we aim to provide a direct comparison of the two techniques for magnetic-field measurements, offering valuable insights into their respective sensitivities and applications.

\section{Experimental setup}

\begin{figure}
	\centering 
 	\includegraphics[width=0.45\textwidth]{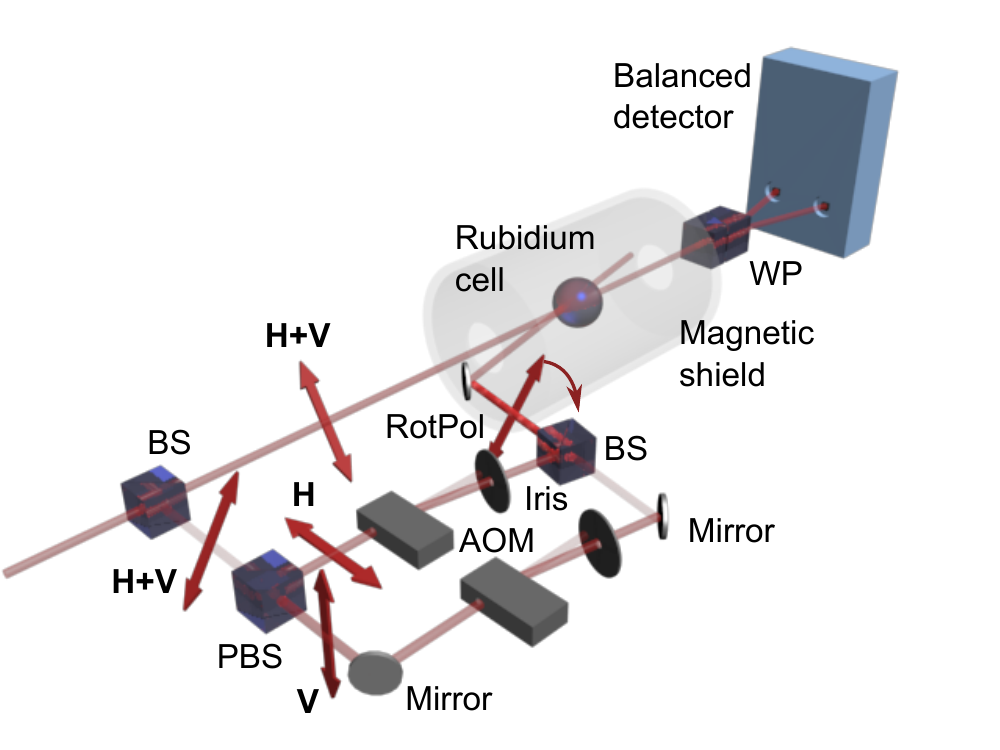}	
 	\caption{Schematic of the experimental setup used for field measurements. BS stands for a beam splitter, PBS is a polarization beam splitter, WP is a Wollaston prism, AOM is an acousto-optic modulator, \textbf{H} and \textbf{V} indicate horizontal and vertical linear polarization, while \textbf{H+V} is the polarization oriented at 45$^\circ$.} 
 	\label{fig:Setup}%
\end{figure}

The scheme of the experimental system used in our measurements is shown in Fig.~\ref{fig:Setup}.
The heart of the system is a spherical vapor cell of 3~cm in diameter, containing an isotopically enriched sample of $^{87}$Rb at a slightly alleviated temperature of 45$^\circ$C. To prolong the atomic polarization lifetime, the cell walls are coated with a special anti-relaxation layer, extending the lifetime to approximately 30\,ms (with polarization preserved in up to 4,000 wall collisions). The cell is placed inside a magnetic shield made of three layers of mu-metal, which enables magnetic-field shielding at a level of 10$^4$. A set of magnetic-field coils is placed inside the shield to compensate for the residual fields and generate a leading field along the probe-light propagation direction.

The cell is illuminated with two light beams: the probe and the pump. If not stated otherwise, both beams are extracted from a diode laser (Toptica DL Pro) with a wavelength tuned to the rubidium D$_1$ line (795\,nm). The linearly polarized light emitted from the laser is split using a beam splitter (BS) and directed into two parts of the experimental setup. The first fraction of the beam is directed into a system generating light with continuously rotating linear polarization. While this system was described in detail in Ref.\,\cite{Wlodarczyk2019System}, here we recall some of its crucial elements. In general, the system is based on a Mach-Zehnder interferometer, in which two orthogonal linear polarizations are directed into separate arms of the device using a polarization beam splitter (PBS). In each arm of the interferometer, an acousto-optic modulator (AOM), optimized for first-order diffraction, is used. Both AOMs are driven with acoustic waves of roughly 80~MHz, extracted from the same generator. While one of the modulators is directly driven by the generator signal, the other driving signal is additionally modulated using a single-sideband modulator, which allows us to coherently shift its frequency by much lower frequency $\nu_m$. The beams directed into the first order of diffraction of both AOMs are then recombined using a beam splitter. The superposition of the orthogonally polarized beams frequency-shifted by $\nu_m$ results in the generation of light with continuously rotating polarization. This light is then used to optically pump the rubidium atoms. It is noteworthy that placing a polarizer after the second BS in the polarization-rotation system allows one to generate amplitude-modulated light. In this way, the RotPol experiment may be combined with the AM NMOR under the same experimental conditions.

The unmodulated part of the original beam is used to illuminate the vapor to probe its spin polarization. After passing through the cell, the polarization of this probe beam is detected using a balanced polarimeter, consisting of a Wollaston prism (WP) that splits the light into orthogonal polarization components and a balanced detector. The difference signal of the detector is proportional to the polarization rotation of light. The balanced-polarimeter output signal is demodulated at the frequency $\nu_m$. This allows the measurements of the NMOR signal versus different experimental parameters such as detuning and both light beams intensities.

A fraction of the probe beam, split out of the probe prior to entering the shield, is used for stabilization and monitoring of the light wavelength. The monitoring is achieved through absorption spectroscopy, while stabilization is implemented using a dichroic atomic vapor laser lock exploiting a microscopic vapor cell \cite{Pustelny2016DAVLL}. In all measurements except those involving wavelength dependence, the light is slightly detuned toward the low-frequency wing of the Doppler-broadened \mbox{$F=2\rightarrow F'=1$} transition.

\section{Results}

\subsection{NMOR signals}

\begin{figure}
    \centering
    \includegraphics[width=0.45\textwidth]{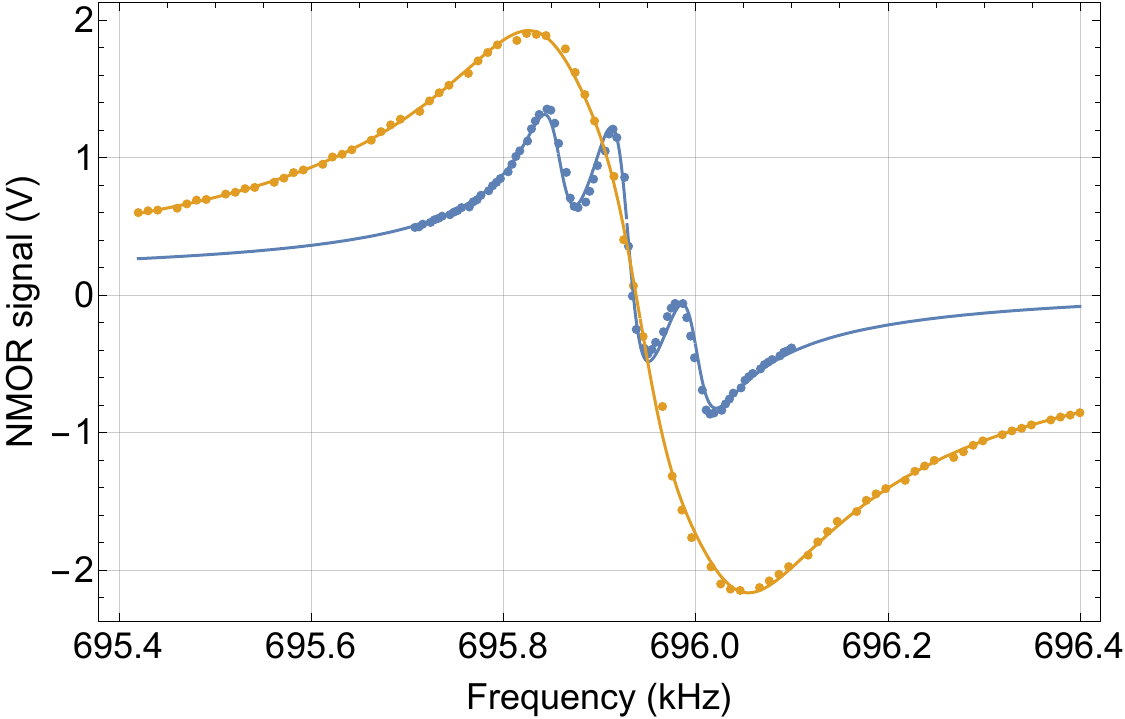}
    \caption{RotPol signals measured at a magnetic field of 100\,$\mu$T versus the polarization-rotation frequency for pump powers of 10\,$\mu$W (blue) and 300\,$\mu$W (orange). The experimental signals (points) are fitted with the triple Lorentzian profile given by Eq.~\eqref{eq:TripleResonance} (solid lines). The signals were measured with a probe power of 10\,$\mu$W.}
    \label{fig:NMOR}
\end{figure}

Figure~\ref{fig:NMOR} shows the signal recorded with RotPol at roughly 100\,$\mu$T, i.e., a field three times larger than Earth's magnetic field in our lab. To investigate the role of the pump-light power, the signals are measured for both weak (10\,$\mu$W) and strong (300\,$\mu$W) pump powers. 
At low pump powers, the observed signal clearly consists of three resonances. This triple structure is a consequence of the nonlinear Zeeman effect, that lifts the degeneracy of the magnetic-sublevel splitting beyond the resonance widths. At higher powers, the amplitudes of the resonances increase, which results from more efficient optical pumping, but the composite resonances are power broadened. In particular, at 300\,$\mu$W, the broadening of individual resonances exceed their width such that the signal manifests as a single resonance. 

To extract quantitative information about the signals, a triple Lorentzian profile is fitted to the experimental data (solid lines in Fig.~\ref{fig:NMOR})
\begin{equation}
    \begin{split}
        f=\bigg(&\frac{A}{\frac{\nu-\nu_0}{\gamma}+i}\\
        +&\frac{A_1}{\frac{\nu-\nu_0+\nu_0^2/\Delta_\text{HF}}{\gamma_1}+i}+\frac{A_1}{\frac{\nu-\nu_0-\nu_0^2/\Delta_\text{HF}}{\gamma_1}+i}\bigg)e^{i\phi},   
    \end{split}
    \label{eq:TripleResonance}
\end{equation}
where $A$ and $A_1$ are the amplitudes of the composite resonances, $\gamma$ and $\gamma_1$ are their widths, $\nu_0$ is the Larmor frequency and hence the position of the central resonance, $\phi$ is the global phase, and $\Delta_\text{HF}$ is the rubidium-87 hyperfine frequency \cite{Steck2010Rubidium}. Note that in the fitting, the satellite resonances have the same amplitude and widths, which, however, are different from those of the central resonance. Fitting the signals enables a comprehensive investigation of the resonance parameters and successively determine the sensitivity of magnetic-field measurements as a function of such experimental parameters as pump and probe powers and tunings. 

\subsection{NMOR with rotating polarization vs amplitude-modulated light}

One of the main questions of this paper is whether the RotPol approach offers superior performance compared to the method utilizing AM light. To investigate this, we measured NMOR signals using two schemes: one employing rotating polarization and the other in which amplitude modulation of pump light is used. When placing the high-quality polarizer after the interferometer, AM light with 100\% modulation at twice the polarization-rotation frequency is achieved, enabling AMOR measurements \cite{Gawlik2006Nonlinear}.

In ensuring a fair comparison between RotPol and AMOR, an important question arises: Should the comparison be based on signals with the same average light intensity or the same amplitude of modulation of the pump (bearing in mind the two-fold difference between the average pump power in these scenarios)? Given the validity of arguments for both cases, we explore and compare a RotPol signal with AMOR signals measured for both cases.

\begin{figure}
    \centering
    \includegraphics[width=0.45\textwidth]{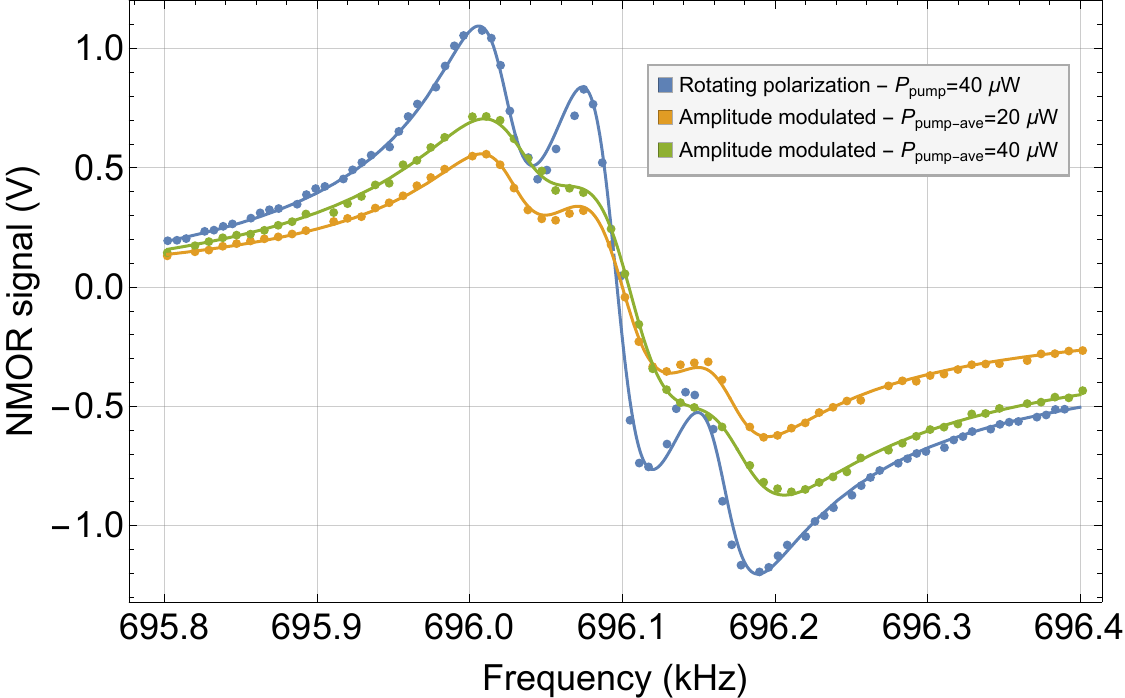}
    \caption{Comparison of NMOR signals measured with RotPol and AM light. The parameters extracted based on triple Lorenzian fits are given in Table\,\ref{tab:AMvsRotPol}.}
    \label{fig:AMvsRotPol}
\end{figure}

Figure~\ref{fig:AMvsRotPol} shows representative RotPol and AMOR signals measured at the same magnetic field (roughly Earth's magnetic field), with identical pump and probe tunings, and the same probe power.
The results highlight a distinct trend: stronger signals are observed when utilizing the rotating-polarization pump. In fact, the RotPol-signal amplitude is roughly two times larger than the AMOR-signal amplitudes (see Table~\ref{tab:AMvsRotPol}). Moreover, in both of the AMOR cases, the widths of the resonances are notably larger. This effect originates from additional relaxation arising due to the pump; as the pump intensity follows a sinusoidal modulation, there is a finite probability of repumping the atoms that were already polarized by light. Since newly created polarization deviates in orientation from the original one, this process deteriorates the overall transverse polarization of the atoms, acting as an additional relaxation. As might be expected, the effect is more pronounced at stronger pump power (Table~\ref{tab:AMvsRotPol}). Specifically while the AMOR-signal amplitude undergoes a change of approximately 25\% when going from 10\,$\mu$W to 20\,$\mu$W, the signals also broaden by about 15\%. In turn, the slope of the resonance, determining the sensitivity of magnetic-field measurements (see below), remains nearly unchanged.

\begin{table}
\centering
\begin{tabular}{l c c c c} 
 \hline
 Type & Pump power & Amplitude & Width & Slope\\ 
    & ($\mu$W) & (V) & (Hz) & (mV/H) \\
 \hline
 RotPol & 10 & 1.947(26) & 27.17(81) & 71.6(3.1) \\ 
 AM & 10 & 0.8952(88) & 35.99(80) & 24.87(80) 	 \\ 
 AM & 20 & 1.1250(82) & 41.68(60) & 26.99(59) 	 \\ 
 \hline
\end{tabular}
\caption{Amplitude, width, and slope of the NMOR signals observed with RotPol and AM light. The AM data are shown for two average light powers: one corresponding to the same peak intensity as in the RotPol case and the other to the same average light power (a factor of 2 difference between the cases).}
\label{tab:AMvsRotPol}
\end{table}

\begin{figure}
    \centering
    \includegraphics[width=0.5\textwidth]{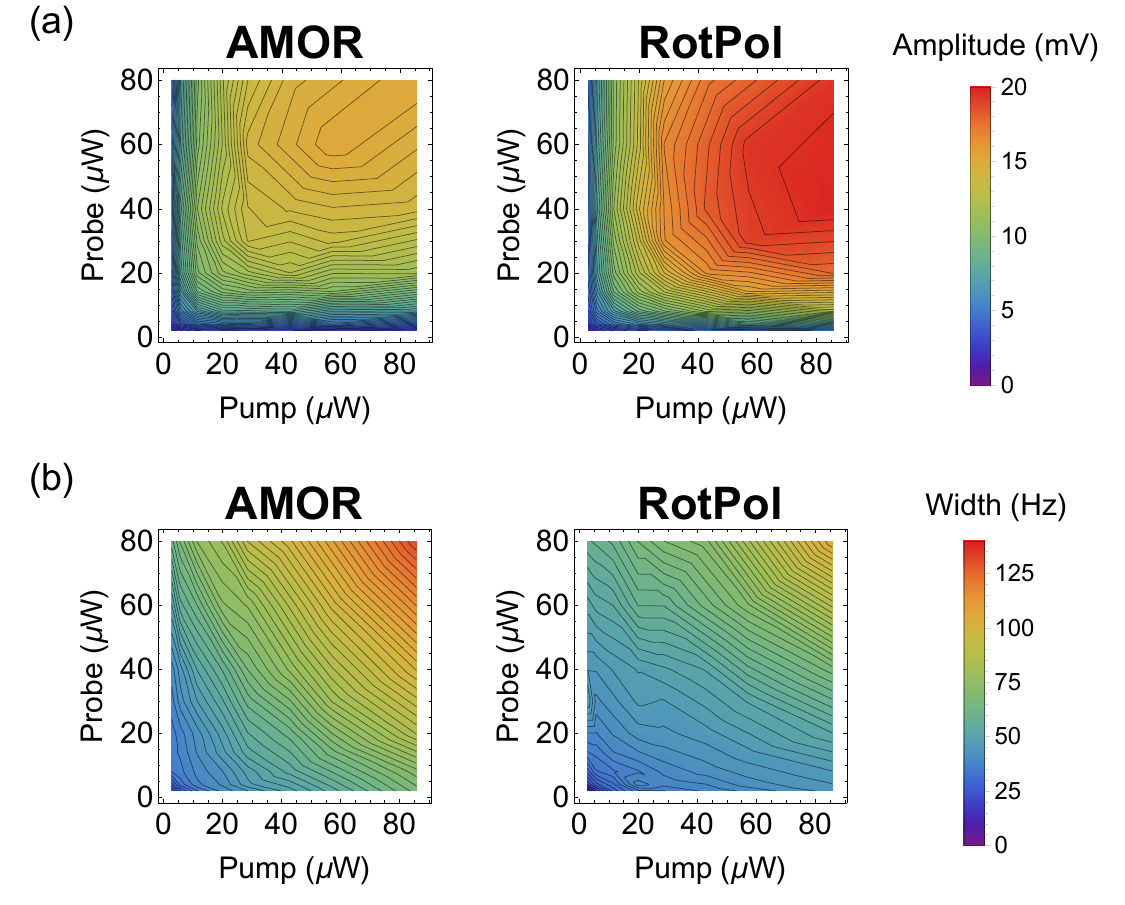}
    \caption{Amplitude (a) and width (b) of AMOR (left column) and RotPol (right column) signals (central resonance) versus the pump and probe powers.}
    \label{fig:AmplitudeAndWidthPumpAndProbe}
\end{figure}

To verify whether under different conditions the efficiency of AMOR generation does not exceed that of the optimum RotPol, we measured the amplitude and width of both signals versus pump and probe powers (Fig.~\ref{fig:AmplitudeAndWidthPumpAndProbe}).
The obtained results show that the amplitude of the resonance measured in RotPol is always larger than that of the corresponding AMOR resonance. Moreover, for RotPol the strongest resonance is observed for relatively strong pump and not-too-strong probe. At the same time, in AMOR, the strongest resonance is observed for weaker pump and stronger probe. The resonance, however, is about 30\% weaker than in the RotPol case. Under the same conditions, the width of the observed NMOR resonances is about 30\% narrower, though the trends in both techniques are the same (the width monotonically increases with the power of either of the beams).

\begin{figure}
    \centering
    \includegraphics[width=0.45\textwidth]{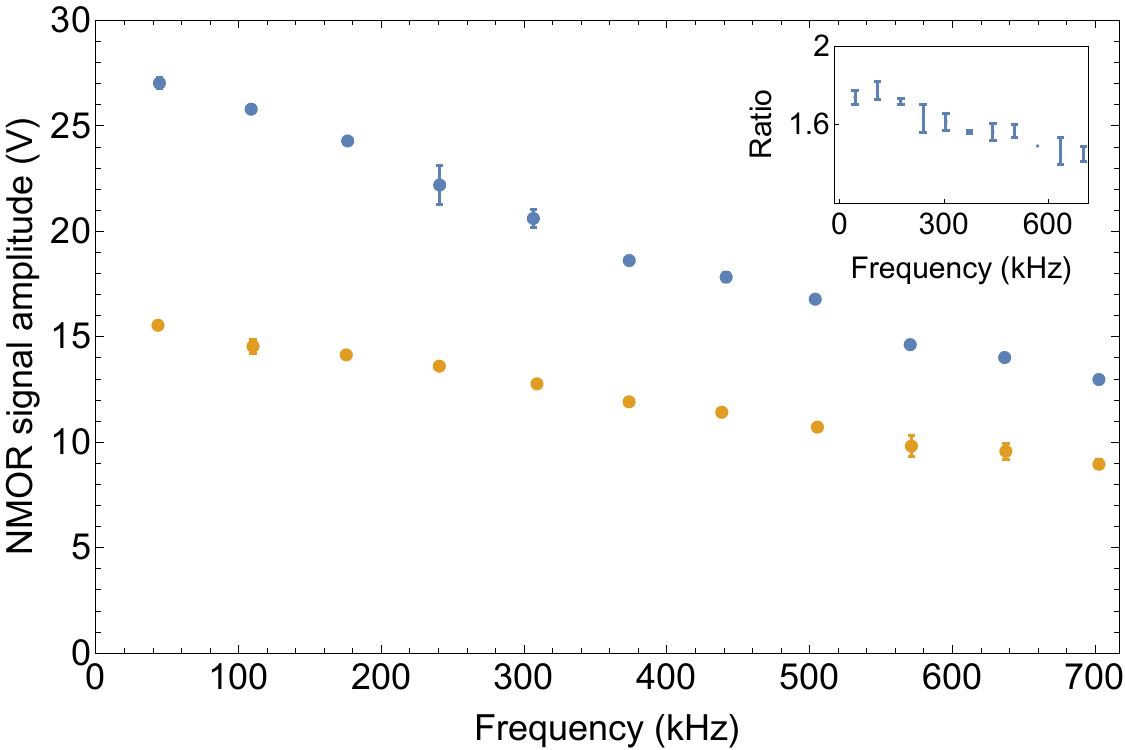}
    \caption{Amplitude of the NMOR signal versus magnetic field measured with RotPol and AM light.}
    \label{fig:MagneticDependence}
\end{figure}

Another interesting aspect of the signals is their dependence on the magnetic field/Larmor frequency. Analysis of this dependence allows one to study the hypothesis of the AOC-induced deterioration of the signal. Figure~\ref{fig:MagneticDependence} shows the amplitude of the NMOR signal measured with both AM and RotPol light versus the magnetic field.
The results show that the amplitudes of RotPol resonances are larger than the AMOR ones (about 1.5 times, as shown in the inset to Fig.~\ref{fig:MagneticDependence}). While the amplitude ratio reveals a relatively weak dependence on the leading field, the individual amplitudes depend on the field, experiencing about a 50\% reduction when transitioning from weak to strong fields. This is somewhat surprising in the scope of previous works, where significantly stronger leading-field dependence was demonstrated \cite{Acosta2008Production}. This also indicates that AOC does not play an important role in the signal deterioration, leaving the question of why such reduction was observed previously. While the answer to the question may only be speculative, one may expect that the effect was either induced by broadening of the resonance due to an increase in the absolute inhomogeneity of the applied magnetic field \cite{Pustelny2006Influence} or by some filtering effect present in experimental systems, leading, for example, to the inability to perform 100\% modulation (where optical pumping was either less efficient or repumping destroyed the atomic polarization). In the presented studies, this effect is not present due to a different technique of generating AM light. 

\subsection{Spectral dependence of the RotPol signal}

To study the spectral dependence of the RotPol signals, the experimental setup was modified by incorporating a separate probe laser (a fraction of the light beam originally used for probing was blocked). This modification allowed for independent control of the tuning of both lasers, facilitating the search for the optimal signal. 

\begin{figure}
    \centering
    \includegraphics[width=0.45\textwidth]{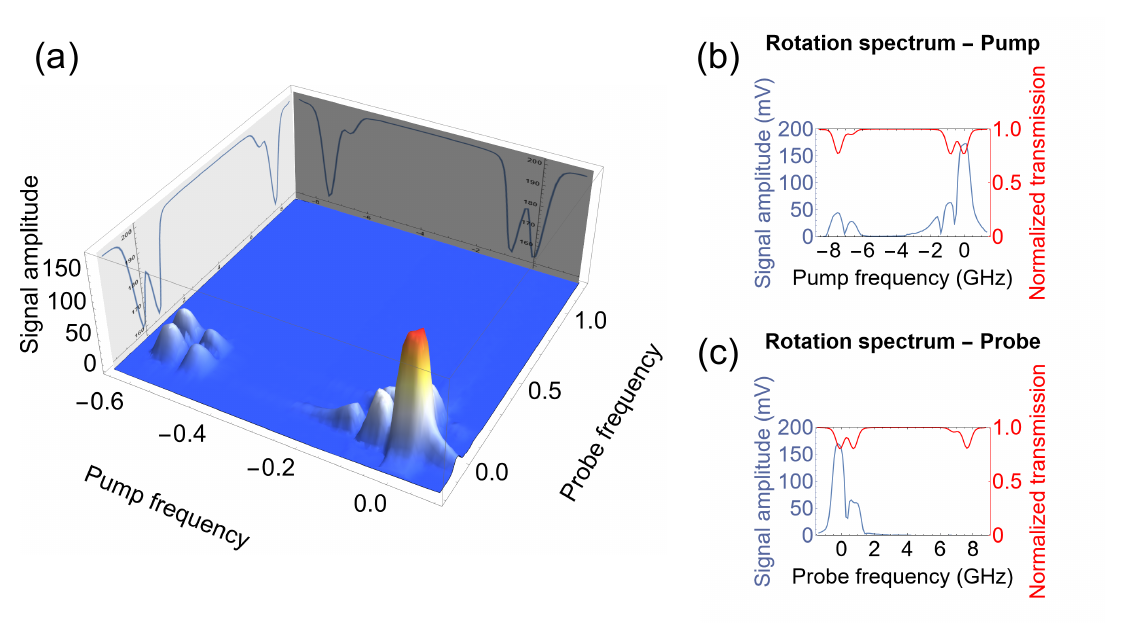}
    \caption{Dependence of the amplitude of the RotPol signal on the tuning of the pump and probe light. Plot (a) shows the signal for a set of various tunings across the D$_1$ line. The two plots on the right, (b) and (c), show projections of the signal into two planes. Plot (b) illustrates the dependence for the probe tuned close to the center of the Doppler-broadened $F=2\rightarrow F'=1$ transition, with the pump scanned across the D$_1$ line. Plot (c) presents the corresponding signal when the pump is tuned to the center of the Doppler-broadened $F=2\rightarrow F'=1$ transition, and the probe is scanned across the same line.}
    \label{fig:WavelengthDependence}
\end{figure}

Figure~\ref{fig:WavelengthDependence} shows the amplitude of the observed signal as a function of the wavelengths of both lasers. 
As demonstrated, the strongest signal is observed when the lasers are tuned to the center of the Doppler-broadened $F=2\rightarrow F'=1$ transition. This suggests that for such tunings, the largest transverse polarization is being generated in the medium and its effect on the probe-light polarization is also the strongest. Weaker signals appear when only one of the lasers is tuned to the \mbox{$F=2\rightarrow F'=1$} transition, while the other laser is tuned to the same ground but different excited state (the $F=2\rightarrow F'=2$ transition). Under such conditions, either the magnitude of the generated atomic polarization or its effect on probe-light polarization is weaker. This originates from the difference in value and opposite in sign dipole matrix elements for the two transitions. In addition, two distinct peaks in rotation are observed for the $F=2\rightarrow F'=2$ pump tuning [Fig.~\ref{fig:WavelengthDependence}(b)]. These peaks mark a reversal in the direction of magneto-optical rotation, marked by the zero crossing (as the plot illustrates the absolute value of the rotation). This phenomenon aligns with behaviors previously observed in various NMOR studies \cite{Budker2000NMOR,Pustelny2006Pump}. Finally, the signal is also observed for both beams tuned to the $F=2\rightarrow F'=2$ transitions, though its amplitude is even smaller compared to previous cases.

The results shown in Fig.~\ref{fig:WavelengthDependence} also indicate the presence of a signal for the pump tuned to the \mbox{$F=1\rightarrow F'$} transition. Although polarization rotation under this tuning is observed in both conventional (unmodulated-light) \cite{Drampyan2008Electromagnetically} and modulated-light NMOR \cite{Budker2002FMNMOR}, the situation in question is qualitatively different. In conventional NMOR, the opposite rotation directions of spins in two ground-state hyperfine levels (nearly opposite gyromagnetic ratios) are irrelevant due to slow precession (the spin relaxation rate is comparable to the Larmor frequency) so that optical pumping of transverse atomic polarization (transverse alignment) is efficient for both tunings. At higher fields, when either FM or AM light is used, the rotation is indeed opposite. In this case, however, the modulated light comprises both spectral components shifted by $\pm\nu_m$, ensuring that one component may be resonant for either tuning (this situation is analogous to nuclear magnetic resonance, where nuclear spins are excited by a component of the oscillating radio-frequency signal that co-rotates with the spins). In the RotPol scenario, where light precesses in a specific direction, the strong dynamic polarization of the medium can only be generated when spins co-precess with light polarization. In turn, in the other hyperfine level, the light does not induce the transverse atomic polarization, but generates an average static atomic polarization (longitudinal alignment), undetectable with polarization rotation of light propagating in the same direction as the pump. Thereby, the rotation signal should only be observed for a given hyperfine tuning. However, our data show that such a  signal is also present for the other transition [Fig.~\ref{fig:WavelengthDependence}(b)]. After careful analysis, it turns out that this signal is not a magnetic-induced signal but rather an artifact originating from the optical pumping. Specifically, the strong pump optically polarizes the atom at any time, creating in this way a weak transverse component of atomic polarization synchronized with its rotation. In this way, the probe experiences modulated properties of the medium, which manifest when the pump is tuned to the other hyperfine state. As this signal does not depend on the magnetic field, it is not interestteresting from the perspective of magnetometry.

Our analysis shows that the optimal conditions for NMOR-signal generation are achieved when the pump and probe have similar tuning. This suggests a possibility of using a single laser for both the pump and the probe.

\subsection{Sensitivity to magnetic fields\label{sec:Sensitivity}}

The results presented above demonstrate that the amplitude $A$ of the NMOR resonance in RotPol is greater than that observed in AMOR and that the width $\gamma$ of the former is narrower than that of the latter. From a practical point of view, the sensitivity of optical magnetometers is determined by these parameters, as well as the signal-to-noise ratio (SNR), as described by the equation:
\begin{equation}
	\delta B=\frac{g\mu_B}{\hbar}\times\frac{A}{\text{SNR}}\times\gamma,
\end{equation}
where $\mu_B$ is the Bohr magneton, $g$ is the Land\'e factor, and $\hbar$ is the reduced Planck constant. In turn, studying the dependence of the parameters on NMOR resonances allows us to compare the sensitivity of RotPol and AMOR magnetometry. 

\begin{figure}
    \centering
    \includegraphics[width=0.5\textwidth]{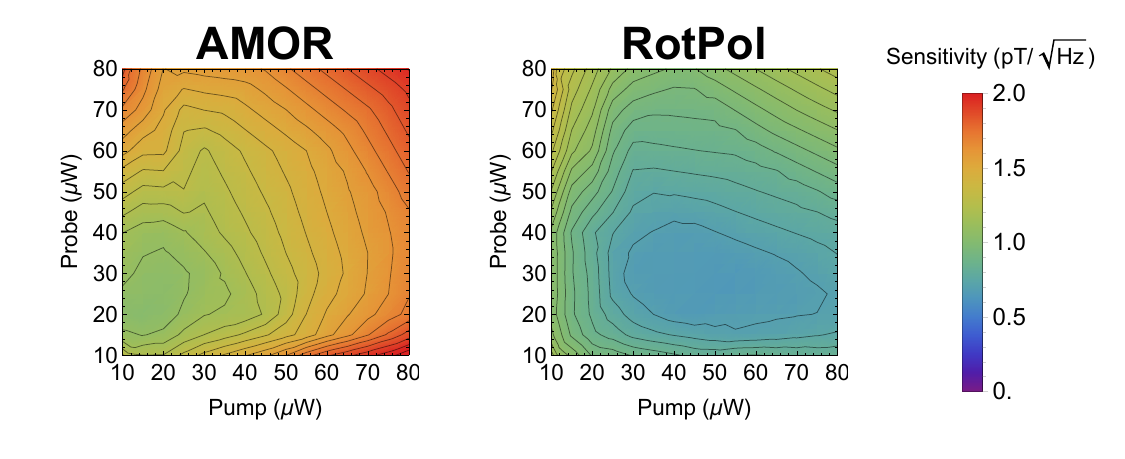}
    \caption{Sensitivity of the stronger ($\approx 30$~$\mu$T) magnetic-field measurements using AMOR (left) and RotPol (right) light versus the pump and probe powers. For the measurements a single laser with a split light beam was used.}
    \label{fig:Sensitivity}
\end{figure}

Figure~\ref{fig:Sensitivity} presents the comparison of RotPol and AMOR magnetometry as a function of pump and probe powers.
The results show that in RotPol, the optimal sensitivity of 650~fT/Hz$^{1/2}$ is achieved with a pump power of 55~$\mu$W and a probe power of about 30~$\mu$W. AMOR achieves an optimal sensitivity of 1.15~pT/Hz$^{1/2}$ with a pump power of 20~$\mu$W and a probe power of 25~$\mu$W. Although the sensitivity might be somewhat lower compared to other techniques \cite{Schwindt2005Self,Higbie2006Robust,Pustelny2008Magnetometry}, it is important to note that these measurements were performed in a magnetic field of 100~$\mu$T, exceeding the typical operating conditions of optical magnetometers by at least a factor of 3, and more often by 3-4 orders of magnitude. This particularly distinguishes this study from other research.

\section{Discussion}

To the best of our knowledge, the presented technique provides the first example of NMOR-based magnetometry applied to fields that significantly exceed Earth's magnetic field (note that unmodulated-light NMOR was previously studied in such or stronger fields \cite{Gawlik1974Strong,GawlikW1987Forward}). Utilizing such an ultrasensitive technique with a broad dynamic range may be important for various applications. For example, in space exploration, NMOR-based magnetometry can be used for the measurement of interplanetary magnetic fields \cite{Owens2013Heliospheric}, as well as fields in celestial and planetary environments, enabling, for example, investigations of Jupiter's magnetosphere \cite{Connerney2017Jupiter} or facilitating the search for magnetic anomalies on Mars \cite{Acuna1999Global}. Moreover, in materials science, examining the magnetic properties of materials under stronger fields may provide new insights into their structure and magnetic or mechanical defects \cite{Ko2022Optically}. Finally, measurements of magnetic fields slightly stronger than Earth's magnetic field can be beneficial in geophysical navigation by detecting magnetic anomalies associated with specific natural resources or facilitating the identification of subsurface geological structures \cite{Akulshin2025Remote}.

A different aspect of the presented research is the potential improvement of the magnetometer performance, e.g., its sensitivity. A specific, yet unexplored aspect of such improvement is the concentration of the medium. As shown in Ref.~\cite{Pustelny2008Magnetometry}, increasing the medium concentration to approximately one optical depth maximizes the sensitivity of magnetic-field measurements. It is clear from Fig.~\ref{fig:Sensitivity} that our magnetometer is far from being optimized in this regard, and it is feasible to increase the concentration by more than an order of magnitude with only minimal compromise in the width of the observed resonance. Adopting this strategy could lead to a ten-fold increase in the sensitivity of magnetometric measurements.

Finally, it is worth noting the potential to use this RotPol technique in a self-oscillation mode \cite{Schwindt2005Self,Higbie2006Robust,Pustelny2007All}. In such a case, the appropriately amplified polarization-rotation signal can be fed back as a modulation signal to the polarization-rotation system. Since the atoms themselves act as a narrow-band filter, this configuration will favor frequencies associated with the Larmor frequency, and spin precession at this frequency will be amplified. This will lead to the precession of spins at the Larmor frequency and successive automatic tracking of the field changes. We have already built such a system and tested its operation \cite{Wlodarczyk2016Podzespoly}.

A challenge with the described solution is developing a simple and reliable system for generating rotating polarization. While a bulk setup using a Mach-Zehnder interferometer combined with acousto-optic modulators is conceptually straightforward, its implementation and reliability pose certain technical issues. Thus, an attractive approach could involve creating such a system using monolithic optoelectronic systems, incorporating the system onto a single chip \cite{Sohler2008Integrated,Boes2018Status}. Integrating an interferometer with phase-shifting elements could potentially address some of the aforementioned technical challenges. Alternatively, electro-optic modulators could be employed to manipulate light polarization. In this case, an electric field would dictate the spatial orientation of the linear polarization. However, the inability to continuously increase the electric field value would necessitate either ``resetting'' the field—causing the light's phase to advance monotonically only within a certain angular range and then abruptly unwrapping the rotation—or developing a solution in which combining electric fields across one or more electro-optic crystals results in continuous light rotation. So far, though, none of these solutions have been fully developed.

\section{Summary and conclusions}

In this article, we presented a novel method for measuring magnetic fields using light with continuously rotating linear polarization. By synchronizing the precession of the polarization with the Larmor frequency of the atomic spins in a medium, a dynamically precessing atomic polarization of significant magnitude is generated within the medium. This transverse anisotropy strongly modulates the properties of the independent, unmodulated linearly polarized light traversing the medium, enabling the detection of nonlinear magneto-optical rotation signals with large amplitudes. This large amplitude polarization signal translates to a high sensitivity for this method compared to other NMOR techniques utilizing modulated light. The method itself demonstrated the capability to measure magnetic fields across a wide range, up to fields nearly three times greater than Earth's magnetic field, as pumping conditions do not depend on light precession frequency. Thus, the method presents a unique combination of very high magnetometric sensitivity—which, as discussed, can be increased by up to threefold—and an exceptionally large dynamic range. This makes our technique potentially useful for precise magnetic field measurements across a broad dynamic range in future applications.

The presented results also allowed us to verify the hypothesis regarding the role of the alignment-to-orientation conversion in the reduction of NMOR signal amplitude with increasing magnetic field strength, as previously observed in the literature. Our study demonstrated that in scenarios where the modulation parameters of the light, including its duty cycle and amplitude, do not degrade with increasing magnetic fields, the observed reduction in amplitude is relatively small (approximately 40\%) and similar for both RotPol and AMOR techniques. This evidence suggests that AOC is not responsible for the amplitude reduction. Instead, the decline is likely due to the nonlinear splitting of the resonances and the gradual decoupling of hyperfine interactions by the magnetic field, as well as the emergence of field inhomogeneities.

\begin{acknowledgments}
The authors thank Dmitry Budker for the stimulating discussions during the stage of technique development and Irena Rodzo\'n{} for the preliminary research on the subject. This work was supported by the National Science Centre of Poland through the Sonata Bis program (grant No. 2019/34/E/ST2/00440).
\end{acknowledgments}

\bibliographystyle{apsrev4-2} 
\bibliography{RotPol} 

\end{document}